\pdfoutput=1
\documentclass[sigconf]{acmart}
\usepackage[shortlabels]{enumitem}
\usepackage{CJKutf8}
\usepackage{xcolor}
\usepackage{hyperref}
\usepackage[normalem]{ulem}
\usepackage{enumitem}

\hypersetup{
    colorlinks=true,
    linkcolor=blue,
    filecolor=magenta,      
    urlcolor=cyan,
    pdfborderstyle={/S/U/W 1}
}

\makeatletter
\begingroup
  \catcode`\$=6 %
  \catcode`\#=12 %
  \gdef\href@split$1#$2#$3\\$4{%
    \hyper@@link{$1}{$2}{\uline{$4}}
    \endgroup
  }%
\endgroup

\newcommand{\cjktext}[1]{\begin{CJK}{UTF8}{}\CJKfamily{mj} {#1}\end{CJK}}

\def\BibTeX{{\rm B\kern-.05em{\sc i\kern-.025em b}\kern-.08emT\kern-.1667em\lower.7ex\hbox{E}\kern-.125emX}}
    
\copyrightyear{2019}
\acmYear{2019}
\setcopyright{acmlicensed}
\acmConference[SIGIR '19]{SIGIR '19: SIGIR Symposium on IR in Practice (SIRIP) 2019}{July 21--25, 2019}{Paris, France}

\begin{document}
%
\title{Challenges in Search on Streaming Services:  Netflix Case Study }

\author{Sudarshan Lamkhede}
\authornote{Both authors contributed equally to this research.}
\email{slamkhede@netflix.com}
\orcid{1234-5678-9012}
\affiliation{%
  \institution{Netflix Inc.}
  \streetaddress{100 Winchester Boulevard}
  \city{Los Gatos}
  \state{California}
  \postcode{95032}
}
\author{Sudeep Das}
\authornotemark[1]
\email{sdas@netflix.com}
\affiliation{%
  \institution{Netflix Inc.}
  \streetaddress{100 Winchester Boulevard}
  \city{Los Gatos}
  \state{California}
  \postcode{95032}
}
\newcommand{\search}{Search}
\newcommand{\instantsearch}{Instant Search}
\newcommand{\recsys}{Recommender Systems}
\newcommand{\userquery}[1]{\textit{\texttt{#1}}}
\newcommand{\movietv}[2]{\href{#2}{#1}}
\newcommand{\suggest}[1]{{\color{red} #1}}
%
\renewcommand{\shortauthors}{Lamkhede and Das}

%
\begin{abstract}
We discuss salient challenges of building a search experience for a streaming media service such as Netflix. We provide an overview of the role of recommendations within the search context to aid content discovery and support searches for unavailable (out-of-catalog) entities. We also stress the importance of keystroke-level \instantsearch{} experience, and the technical challenges associated with implementing it across different devices and languages for a global audience.      
\end{abstract}


%

\maketitle
\section{Introduction}
Streaming media services such as Spotify, iTunes and Pandora etc. for music, and YouTube, Netflix, Amazon Prime Instant Video, HULU, and HBO etc. for video have witnessed large scale adoption in recent years. Convenience, control and choice offered by these services have made them indispensable in broadband connected households \cite{LRG}. 

Most streaming platforms provide users with access to
vast repositories of content with only a small fraction familiar to them. \recsys{} typically do the heavy lifting of providing the user with a relevant subset of items, e.g. the Home Page experience on Netflix, or the personalized Daily Mix recommendations on Spotify. However, \search{} on these services is critical for discovery and exploration of content. In particular, \search{} is the main avenue for new users to explore the content catalog, and for tenured users to break out of the so-called filter bubble \cite{1902.10730}. On Netflix, too, the Recommendation System is the primary driver of discovery (act of watching content that was not watched at all by a user on our service, previously), and \search{} plays a complementary role to the personalized recommendations. We have observed that more than 20\% of discovery streaming happens through \search{} on Netflix. Users have different use-cases for search on streaming media platforms that go beyond traditional information retrieval and require seamless integration of recommendations into search results. For instance, searches for videos that are not available for streaming (``unavailable entities'') require relevant recommendations to be surfaced that are related to the unavailable entity. Also, unlike Web Search, a lot of  search activity happens on disparate devices - smartphones, TVs, game consoles etc. posing unique problems of device adaptation. All these novel interactions lead to \search{} on streaming services being significantly different from Web Search or traditional IR systems.  
\section{\search{} use cases: fetch, find, and explore}
\begin{figure*}[h]
    \includegraphics[width=1.7\columnwidth]{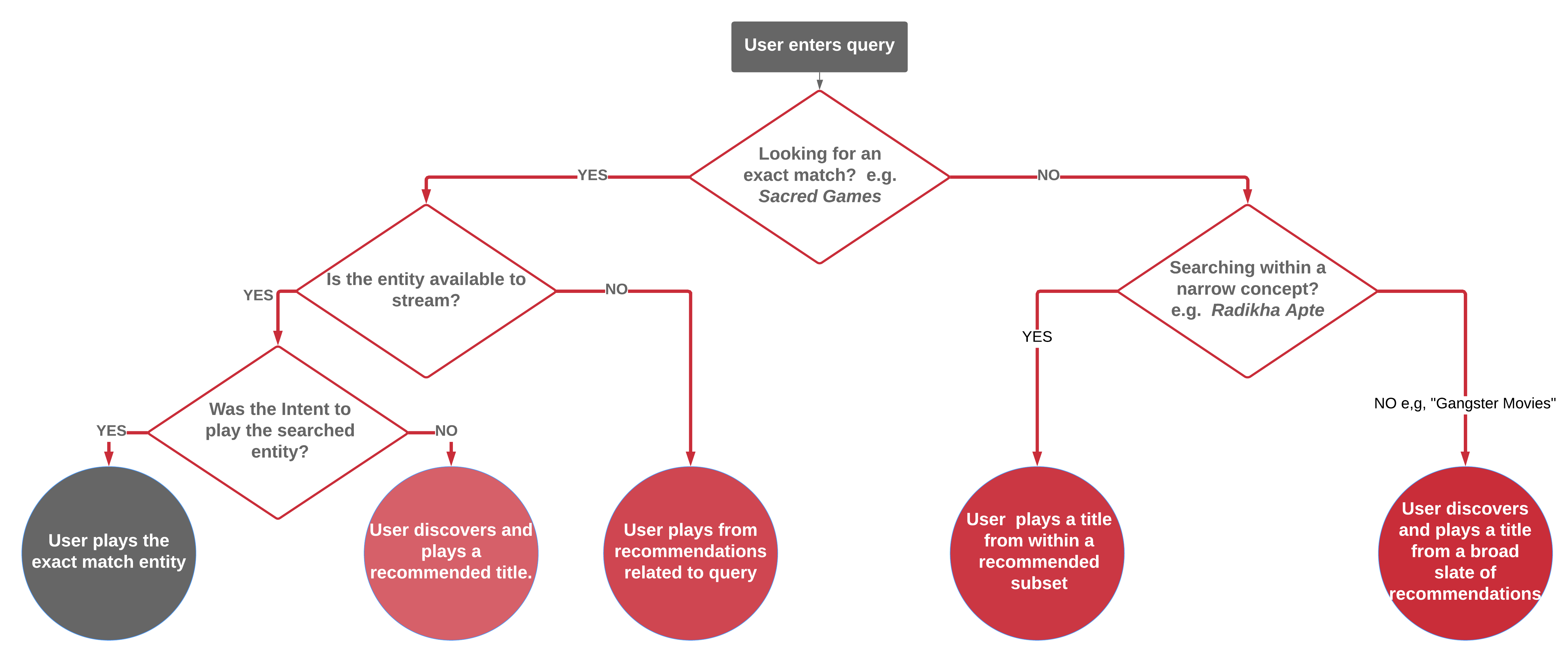}
    \caption{Possible paths connecting a query to a successful play event, illustrating the different use-cases of \search{} on Netflix. Non-lexical, behavioral recommendations become increasingly important for the terminal nodes shown from left to right. The leftmost terminal node is the only one that can be supported by purely lexical information retrieval.}
    \label{fig:search_to_play}
\end{figure*}
User studies at Netflix have revealed three different mindsets in which members interact with \search, namely, \textit{Fetch}, \textit{Find} and \textit{Explore}, even though the aim is to watch something for entertainment. This is quite different than the intents of navigation, information, or transaction behind the Web \search{} queries \cite{Broder:2002:TWS:792550.792552}. 

\noindent\textbf{Fetch:} In this use-case, the users have a clear intent of retrieving a specific item from the catalog to stream. For example, a user who wants to watch \movietv{Stranger Things}{https://www.imdb.com/title/tt4574334/} may issue a query \userquery{stranger things}. In this case users want the system to immediately satisfy their needs by returning the entity.  This is by far the most common use case and largely satisfied by traditional information retrieval techniques that rely on lexical matching.

\noindent\textbf{Find:} In this use-case, users have formulated their entertainment needs but they do not have a specific item in mind. An example search would be for \userquery{radhika apte}. The actress stars in multiple movies and TV shows and the user may be willing to watch any of those. The users' expectation is that the query is understood and relevant items are returned. They have the perception of a partnership between the them and the \search{} system that will enable them to narrow down the choice to a single video.

\noindent\textbf{Explore:} In this use-case, users typically enter much broader queries, such as \userquery{horror movies} or \userquery{spanish} and the idea is to explore the content in that general area. The role of the \search{} system is to provide a slate of relevant results and guide the user through a meaningful journey to a title they would finally choose to watch. 

While discovery of content is the main goal of the explore use-case, it can happen in other use-cases too. Only the fetch use-case is supported by a simple lexical match. Therefore, \search{} needs to blend in both lexical and behavioral results to provide a richer, more meaningful experience. 

\section{Role of recommendations in the search context}

We define a \textbf{Search Result} as an entity retrieved by the search engine by matching query and context with the indexed entities. e.g. results from lexical matches. A \textbf{Recommendation}, on the other hand, is an entity selected by the search engine by relaxing the match constraints e.g. an entity retrieved via collaborative filtering. 

The \search{} system has to retrieve the items that satisfy the query intent in the narrow sense (e.g. return all available James Bond movies if the query is \userquery{james bond}). Additionally, it has to delight users by helping them discover something that they would like to stream. Though the majority use-case (roughly 75\% of searches) for searching is  fetching videos by their titles, a significant portion of users engage in a more relaxed searching behavior. 
e.g. certain fraction of users would discover (and play) \movietv{Piranha movies}{https://en.wikipedia.org/wiki/Piranha_(film_series)} when they searched for \texttt{sharks} with a possible intent of watching  \movietv{Jaws}{https://www.imdb.com/title/tt0073195/}. Some users like to co-search and co-watch videos for Bruce Lee and Jackie Chan. These intents are better served by recommendations.

In case of streaming music services, user studies have shown that serendipitous discovery is highly valued by listeners \cite{Lee:2016:UMI:3039290.3039293} and can increase their long term satisfaction with the service leading to continued subscription \cite{matti}. We hypothesize that those observations hold true for video streaming as well.
 
\par Providing meaningful recommendations within a \search{} experience poses two main technical problems: 
\begin{enumerate}[wide, labelwidth=!, labelindent=0pt,topsep=0pt]
\item  \textbf{What are good recommendations for a search?} - The user query and the search context put restrictions on what could be relevant recommendations. The query agnostic default personalized ranking that we show on the home screen, or unpersonalized, popular entities are not good recommendations in the search context. Such results could be too distracting and may not increase users' engagement or streaming activity.\par So the recommendations need to be contextual and related to the intent.
    For example, \movietv{Game of Thrones}{https://www.imdb.com/title/tt0944947} and \movietv{Breaking Bad}{https://www.imdb.com/title/tt0903747} may not look similar based on the knowledge graph but both are highly binge-worthy shows\footnote{Shows that users may like to watch multiple episodes of, in rapid succession}. So a user that is looking for the next binge-worthy show to watch, either would do. The relevance is also temporal. For example consider query \userquery{oscar nominees}. The intended result set is an ephemeral collection that groups disparate videos prior to the award announcements. We also want the recommendations to be personalized to some extent. When a user seeks related videos for \movietv{The Mummy (1999)}{https://www.imdb.com/title/tt0120616/} movie, they may tend to emphasize the ``action-adventure'' aspect over ``horror'', or ``depicted-era'' over ``cast'' in determining what is related. So we have to devise new ways of coming up with recommendations in presence of the extra information. Traditional query expansion techniques or pure co-play based result set augmentation are not adequate.
    
\item \textbf{How to blend the recommendations with search results?} This can be cast as a re-ranking problem: we can get two sets of videos - search results and recommendations (as defined above) and have another ranking function to rank the combined set. Such blending introduces specific  challenges: 
    \begin{enumerate}[wide, labelwidth=!, labelindent=0pt,topsep=0pt]
        \item The final ranking should respect that strong lexical matches be surfaced at prominent positions when the intent is to retrieve that item  e.g. we should be able to show a documentary \movietv{Shark}{https://www.imdb.com/title/tt4706496/} for query \userquery{shark} at a prominent position, and not bury it within behaviorally relevant entities. We expect \movietv{Black Panther}{https://www.imdb.com/title/tt1825683} to be prominently placed in results when users query it by a prefix of the title.
        \item  The final ranked result set should look relevant overall without any obvious quality issues or inadvertent sensitive or offensive entities that make users feel like the system let them down. 
    \end{enumerate}  
\end{enumerate}

\subsection{Unavailable Entities}
The set of videos available to stream in a market (usually defined at a country level) is limited due to licensing requirements and business constraints. This problem of uneven video availability is described from \recsys{} perspective in \cite{world}. Even the tenured users are not aware of these restrictions. They often search for shows or movies that are not available for streaming. We estimate that at least 13\% of searches on Netflix are for out-of-catalog videos. We need to detect their desired intent and entity, and if the entity is not available for the users to stream, we have to provide them with relevant and delightful recommendations that the users are likely to stream in absence of the original entity. 

This problem arises in E-Commerce setting as well \cite{Singh:2011:UBZ:2009916.2009930, Singh:2012:RNE:2187980.2187989} (e.g. a retailer may not carry a particular brand but not other) but is not noticeable for Web Search backed by a comprehensive crawled corpus spanning the entire Web. 

We do not want the users to hit a dead-end in their efforts to find something worthy of streaming if they issue an unavailable entity search. This goes beyond just plainly recommending related or similar entities. And personalization can play a key role here. 

\noindent There are three distinct sub-problems here: 
\begin{enumerate}[wide, labelwidth=!, labelindent=0pt,topsep=0pt]
    \item \emph{How to detect that the query is for an unavailable entity?} We could index all known entities in the domain of movies and TV shows and then match the user query against those. Keeping such knowledge base up to date and accurate in all languages is very resource consuming. Further, the query could match available and unavailable entities simultaneously and/or match multiple unavailable entities. In such cases, narrowing the intent becomes harder.
    \item \emph{Whether to and how to message the user about unavailability?} Even when we unambiguously know which entity user queried for, messaging it back to the user is nontrivial from the UI perspective. Additionally, it is unclear what effect the messaging would have on the users' perception of the service. 
    \item \emph{How to provide substitutable entities} i.e. entities that can also fulfill the broader intent but aren't exactly the requested entity? Unavailable entities are not present in the co-play data so the traditional collaborative filtering models can not derive item-item similarity for them. Pure metadata based similarity leads to sub-optimal user experience as it fails to account for user behavior patterns. Also, depending on the searched entity, it is possible that users would not stick to their original intent if the  entity unavailable.
\end{enumerate} 

\section{Instant Search}

For long-form video entertainment, users are presumably in a laid-back consumption mode. With least amount of interaction and cognitive effort, they would like to satisfy their entertainment need. Also, most of the viewing happens on TVs. Unlike handheld devices, their on-screen keyboards (OSK) are hard to use with remote controls or pointers. To type a single character, multiple movements of the cursor maybe required on the TV OSK. Voice search isn't that ubiquitous yet and second-screen experiences are not seamless. So to reduce the need to enter multiple characters, the Netflix service offers ``Instant Search'' meaning with every keystroke we provide a set of useful results, instantly.

\instantsearch{} was found to have higher success rate and lower time-to-find in a separate analysis of query logs \cite{DBLP:conf/webdb/CetindilELN12}. If users are not sure about their information need, e.g., if they do not know the correct spelling of the name they are searching for, they will probably make mistakes during typing. Similar to auto complete or query suggestions, \instantsearch{} can guide users along the typing process allowing them to notice and correct mistakes as quickly as possible so that they need to enter as few keystrokes as possible to get to desired result(s). 

While instant search is very helpful for users, it introduces some technical problems. First, it makes the queries very short, second, it makes latency requirements stricter and third, different metrics and indexing schemes are required to optimize the experience.

\subsection{Short Queries}
 The median number of tokens in query in Netflix's query logs is just 1 (table \ref{tab:qlen}). In comparison, the average length of typed queries on Web Search was found to be 2.35 terms \cite{Silverstein:1999:AVL:331403.331405}. The queries on Netflix \search{} are short in number of characters they contain, too. They are even shorter on the TVs - median length is just 3 characters! This makes it difficult to use traditional approaches for query understanding and rewriting. One such example is spell correction. Due to partial queries, it is harder to detect whether the query is misspelled or just incomplete and offer appropriate corrections automatically. The differences in median query length on different platforms / UIs are due to the relative ease of typing. TV UI is the hardest so queries are shorter while Web UI on computers is the easiest so users type longer queries on it.

\begin{table}
  \caption{Query Length on Different Devices: Tokens (Characters). Instant Search leads to very short queries}
  \label{tab:qlen}
  \begin{tabular}{cccccl}
    \toprule
    Platforms / UIs & 1\% & 25\% & 50\% & 75\% & 99\% \\ 
    \midrule
    Android OS & 1 (1) & 1 (4) & 1 (5) & 2 (8) & 5 (23) \\
    iOS & 1 (1) & 1 (4)	& 1 (5) & 2 (8) & 4 (20) \\
    TV UI & 1 (1) & 1 (2) & 1 (3) &	1 (5) & 3 (15) \\
    Web	UI & 1 (2) & 1 (4) & 1 (6) & 2 (9) & 4 (21) \\
  \bottomrule
\end{tabular}
\end{table}

\subsection{Latency}

Users are more likely to perform clicks on the result page that is served with lower latency according to the large scale query log analysis described in \cite{Arapakis:2014:IRL:2600428.2609627}. This is probably true for all search services. Further, \instantsearch{} results need to be rendered almost as soon as user enters a keystrokes. If the results are not instantaneous, user experience degrades, lowering user satisfaction. This requirement puts even stricter latency constraints - both on the UI and the back end - compared to other forms of search. Under these constraints, it becomes crucial to save every millisecond of computation which severely limits design of query understanding, retrieval and ranking components. We have to perform the computation, as much as possible, apriori and rely on efficient and clever caching schemes at serving time. The UI has to render the slate of result without any flickering or jarring updates to the view. Since UI is making multiple requests for a single search, there is more potential for timeouts and errors especially over unreliable network connections.

\subsection{Metrics and Indexing}
One of the objective of offering \instantsearch{} is to help users find what they want to watch with least number of interactions. So driving down number of \emph{keystrokes} to get a desired video in the view port is important, even though it leads to shorter queries that are harder to understand. This is different than traditional search ranking metrics which do not take into account number of interaction required to enter the query. Note that higher number of keystrokes may be required to type a query than the number of characters it has. This can happen due to query reformulation or the language's input method may require it. For example, Korean is usually typed using the Hangul alphabet where syllables are composed from individual characters. For example, to search for 
\cjktext{올드보이} (Oldboy), in the worst possible case, a member would have to enter nine characters: \cjktext{ㅇ ㅗ ㄹㄷ ㅡ ㅂ ㅗ ㅇㅣ}. Using a basic indexing for the video title, in the best case a member would still need to type three characters: \cjktext{ㅇ ㅗ ㄹ}, which would be collapsed in the first syllable of that title: \cjktext{올}. In a Hangul-specific indexing, a member would need to write as little as one character: \cjktext{ㅇ}.  

\section{Global Audience and Content}

\begin{figure}[h]
    \includegraphics[width=0.9\columnwidth]{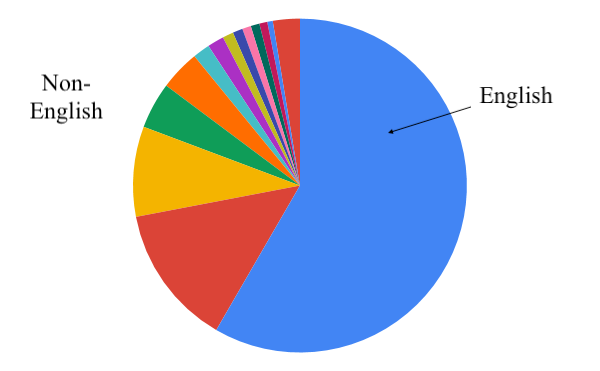}
    \caption{More than 40\% of users search in a language other than English}
    \label{fig:user-languages-in-search}
\end{figure}

Many streaming services are global e.g. Spotify is available in 78 countries. Netflix is global - roughly 60\% of Netflix’s members are outside US and and a significant minority do not consume content in English at all. Netflix is localized in 22 languages and that list is growing. Though English is the top language for users on Netflix \search{}, less than 59\% of users use it for searching (figure \ref{fig:user-languages-in-search}). This proportion is likely to go down as we localize our service in other languages and continue to grow internationally.

We have to offer Search that operates well across all languages, countries and regions. In that sense, the challenges are similar to those in Web Search which also needs to tune the experience in each market. However, for \recsys{} and \search{} on Netflix the differences in country specific corpora and local languages are more fundamental. Each market greatly varies in inclination towards local vs. international content, cultural tastes, query patterns, and content availability. Nonetheless, we are looking into transfer learning from US English to other locales.

How Netflix on boards a new language in \search{} is well described in the blog post \cite{global}.
Though our \search{} works on semi-structured documents that are relatively cleaner and smaller compared to the crawled documents from the Web or e-Commerce product feeds, the localization of the content in all supported languages is usually a challenge. A large number of movies and TV series are released every week all over the world. Keeping the knowledge base updated with all the released and yet-to-be-released entities across the globe with appropriate localization of text while maintaining the knowledge base integrity and high quality is expensive.

Localization sometimes poses an interesting challenge. While localized titles are easy to understand in the target market/locale, their original title may become so popular that users search them by the original title (e.g. \movietv{La Casa De Papel}{https://www.imdb.com/title/tt6468322/} was localized as \textit{Money Heist} in English) causing a mismatch between the indexed string, title image and user query.

\section{Discussion}
This paper describes how the unique user expectations from   \search{}  on a streaming media platform warrant approaches that need to go beyond traditional information retrieval and lean more toward behavioral data.  \search{} and \recsys{} need to work hand-in-hand to increase user joy via content discovery. Many of the challenges stem from the users' intent to play an entity that may be unavailable to stream, or their desire to explore the catalog via \search{}, the limitations of input devices prompting shorter queries, as well as the multi-lingual aspect of search for a global audience. While the specific task, user interface, and user interaction mode may differ between services, we believe that these challenges are relevant for \search{} on all streaming platforms. We hope that the novelty and  practical importance of these problems will attract researchers, both in industry as well as in academia.

\begin{acks}
We are thankful to Jon Sanders, Aish Fenton, Yves Raimond, and Justin Basilico, for their helpful feedback. We are grateful to Priya Kothari for the invaluable user studies narrative and to Yves Raimond for the example of keystrokes matching in Hangul locale.
\end{acks}

\bibliographystyle{ACM-Reference-Format}
\bibliography{sample-base}

\end{document}